\documentclass[pdflatex,sn-mathphys-num]{sn-jnl}
\usepackage{graphicx}%
\usepackage{multirow}%
\usepackage{amsmath,amssymb,amsfonts}
\usepackage{amsthm}%
\usepackage{mathrsfs}%
\usepackage[title]{appendix}%
\usepackage{xcolor}%
\usepackage{textcomp}%
\usepackage{manyfoot}%
\usepackage{booktabs}%
\usepackage{algorithm}%
\usepackage{algorithmicx}%
\usepackage{algpseudocode}%
\usepackage{listings}%
\usepackage[left]{lineno} 

\theoremstyle{thmstyleone}%

\theoremstyle{thmstyletwo}%

\theoremstyle{thmstylethree}%

\begin{document}

\title[Quantum Mechanics of an Abrikosov Vortex in Nanofabricated Pinning Potential]{Quantum Mechanics of an Abrikosov Vortex in Nanofabricated Pinning Potential}

\author[1]{\fnm{Elmeri O.} \sur{Rivasto}}\email{elmeri.rivasto@protonmail.com}

\affil[1]{\orgdiv{Department of Physics, Chemistry and Pharmacy}, \orgname{University of Southern Denmark}, \orgaddress{\street{Campusvej 55}, \city{Odense}, \postcode{5230 Odense}, \country{Denmark}}}

\abstract{A superconducting device is proposed for experimentally investigating whether an Abrikosov vortex can be modeled as a quantum mechanical quasiparticle. The design process of a type-II superconducting device capable of reliably pinning a single Abrikosov vortex is presented, creating a particle-in-a-box-like system. The proposed device consists of a cylindrically symmetric Nb film, 30\,nm in diameter and 5\,nm thick, with a 14\,nm diameter artificial pinning center at its center. Time-dependent Ginzburg-Landau simulations indicate robust single-vortex pinning under an applied field of 6\,T. The presumed quantized energy levels and associated quantum wavefunctions of the vortex quasiparticle are obtained by numerically solving the two-dimensional time-independent Schrödinger equation for this system. It is shown that distinguishing the ground and first excited states is experimentally feasible. Beyond fundamental physics studies, the application of the proposed device in cryogenic memory technology and quantum computing warrant further exploration.}


\maketitle

\section{Introduction}
\noindent Abrikosov vortices are quantized units of magnetic flux that emerge in the mixed state ($B_\mathrm{c1}<B<B_\mathrm{c2}$) of type-II superconductors. They can be modeled as localized degradations in the superconducting order parameter through which a single unit of the magnetic flux quantum ($\Phi_0 = 2.07\cdot 10^{-15}\,\mathrm{Wb}$) can penetrate \cite{Blatter1994vortices}. The (classical) dynamics of the vortices is mainly governed by three distinct interactions: i) repulsive vortex-vortex interaction, ii) attractive interaction (pinning) between a vortex and inhomogeneities in the superconducting lattice and iii) Lorentz force induced by applied current through the superconducting lattice. Since the free movement of vortices due to the Lorentz force results in power dissipation, vortex pinning has been widely studied due to its relevance, in particular, in power transmission and magnet applications of high-temperature superconducting materials \cite{Matsushita2007flux, Foltyn2007materials, Coombs2024high, Shimoyama2024current, Wang2024review}. While the existence of Abrikosov vortices in superconducting films has been generally considered disadvantageous, their true potential in delicate precision electronics has been only recently recognized. Most notably, Abrikosov vortices have been proposed as classical bits in cryogenic memory cells with successful proof of concept experiments \cite{Ortlepp2014access, Golod2015single, Golod2023word, Alam2023cryogenic}. Advancements in possibilities to manipulate and measure single Abrikosov vortices have also shown potential in quantum technology related applications \cite{Togawa2005direct, Rydh2009field, Golod2010detection, Veshchunov2016optical, Keren2023chip, Foltyn2024quantum}. 

Despite being fundamentally quantum objects, the dynamics of the Abrikosov vortices is widely considered to be fully classical due to strong viscous damping resulting from quasiparticle excitations in the vortex core \cite{Bardeen1965theory, Blatter1994vortices}. Classical dynamics has indeed been repeatedly concluded in magnetic force microscopy (MFM) measurements performed for single Abrikosov vortices in high-temperature superconducting thin films \cite{Moser1998low, Auslaender2009mechanics, Brandt2010deforming, Brandt2010nanomechanics}. Novel experiments relying on superconducting quantum interference devices (SQUIDs) have come to the same conclusion \cite{Park1992vortex, Embon2015probing, Gardner2002manipulation, Breitwisch2000pinning}. However, hints for collective quantum tunneling of Abrikosov vortices (quantum creep) in magnetic relaxation experiments have been widely reported \cite{Mota1992quantum, Blatter1994vortices, Prost1993quantum, Beauchamp1995vortex, Seidler1995low}, but these experimental observations have been reasonably questioned by classical mirco-jumps \cite{Miu2008origin} and self-heating effects \cite{Gerber1993self, Eley2021challenges}. However, all of the above-cited experiments consider free or weakly pinned vortices, for which highly dissipative classical dynamics can be expected. In contrast, for strongly pinned vortices whose cores are fully confined within large-diameter columnar defects, where the superconducting order parameter is already suppressed prior to vortex entry, the influence of quasiparticle excitations in the vortex core, and the resulting viscous damping, can be considered highly suppressed. Therefore, quantum-governed dynamics of strongly pinned Abrikosov vortices appear plausible.

While the quantum dynamics of Abrikosov vortices remains debated, quantum effects have been widely observed for Josephson vortices \cite{Ustinov1998solitons} in Josephson junction arrays \cite{Fazio2001quantum}. The quantum dynamics of Josephson vortices was ultimately confirmed by pioneering experiment by Wallraff~\textit{et al.}~\cite{Wallraff2003quantum}, who managed to directly measure the quantized energy levels and quantum tunneling of a single Josephson vortex in Nb/AlO$_x$/Nb junction below 1\,K temperature. Meanwhile, the only direct quantum mechanical effect associated with Abrikosov vortices considers quasiparticle bound states (Caroli--de Gennes--Matricon~(CdGM)~states) in the non-superconducting core of the vortex \cite{Caroli1964bound, Hess1989scanning, Chen2020observation}. It is important to distinguish the CdGM states from the herein considered dynamical collective vortex–quasiparticle states, as they are unrelated to each other. The spacing of the CdGM states is distorted by the presence of point-like defects in the superconducting lattice, giving rise to the elementary pinning force \cite{Balatsky2006impurity, Han2000effect}. This microscopic pinning interaction has been recently confirmed by Chen~\textit{et~al.}~\cite{Chen2024revealing} using high-resolution scanning tunneling spectroscopy. While the quantum description of pinning \cite{Han2000effect} is valid for weak point-like pinning centers, such as atomic vacancies, vortex pinning in nm-scale artificial defects \cite{Zhang2022progress} must still be addressed using the traditional Ginzburg-Landau theory based approaches \cite{Blatter1994vortices, Matsushita2007flux}. 

In this work we present the rigorous process of designing a superconducting device that pins a single Abrikosov vortex, and show how the experimental realization of such a device could be used to determine whether or not Abrikosov vortices behave like quantum mechanical quasiparticles. The manuscript is organized as follows; in section~\ref{designing_the_device_Section} we present detailed discussion on the choice of material (sec.~\ref{material_section}) together with device geometry and dimensions (sec.~\ref{dimensions_section}). In section~\ref{Pinned_Abrikosov_vortices_section} we model the pinning potential of the Abrikosov vortex in the device (sec.~\ref{pinning_potential_section}) and discuss about its effective mass (sec.~\ref{mass_of_abrikosov_vortex_section}). In section~\ref{quantization_section}, based on the conclusions of the previous sections, we then numerically solve the energy levels and quantum wavefunctions of the pinned vortex from a two-dimensional time-independent Schrödinger equation. Finally, we discuss the experimental readout of the quantum state of the vortex in section~\ref{measuring_quantum_state_section} after summarizing the main conclusions in section~\ref{conclusions_section}.

\section{Designing the device}
\label{designing_the_device_Section}
\begin{figure}[t!]
\begin{center}
\includegraphics[width=6cm]{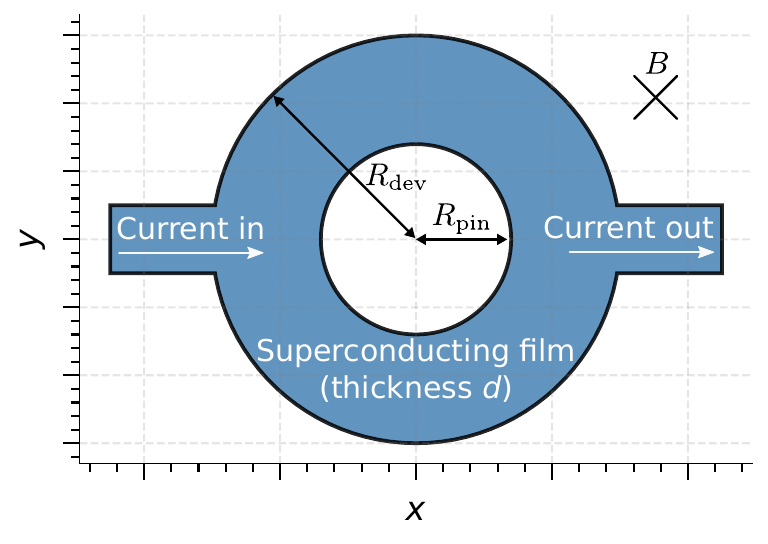}
\caption{A schematic illustration of the design of the Abrikosov vortex qubit device. In this work, we consider the optimization of the superconducting material along with the film thickness (d), the radius of the artificial pinning center ($R_\mathrm{pin}$), radius of the superconducting island ($R_\mathrm{dev}$) and the applied field ($B$) that enables single vortex entry and pinning in the proposed design. }
\label{device_general_illustration_fig}
\end{center}
\end{figure}

\noindent The general design of the proposed device is a simple superconducting thin film of thickness $d$, patterned into a round island of radius $R_\mathrm{dev}$. A round pinning center of radius $R_\mathrm{pin}$ is etched in the center of this island, where we strive to trap the single vortex. Two current terminals of widths $2R_\mathrm{dev}/3$ are also included for manual desolation of the pinned vortex and for control and readout of the device. This design and associated length scales are illustrated in Fig.~\ref{device_general_illustration_fig}. In this section, we will address the choice of superconducting material (and substrate), film thickness and device dimensions $R_\mathrm{dev}$ and $R_\mathrm{pin}$, together with the magnitude of the applied magnetic field (perpendicular to the surface of the film) required to trap a single vortex in the film under zero applied current.

\subsection{Material}
\label{material_section}
\begin{figure*}[t!]
\begin{center}
\includegraphics[width=6cm]{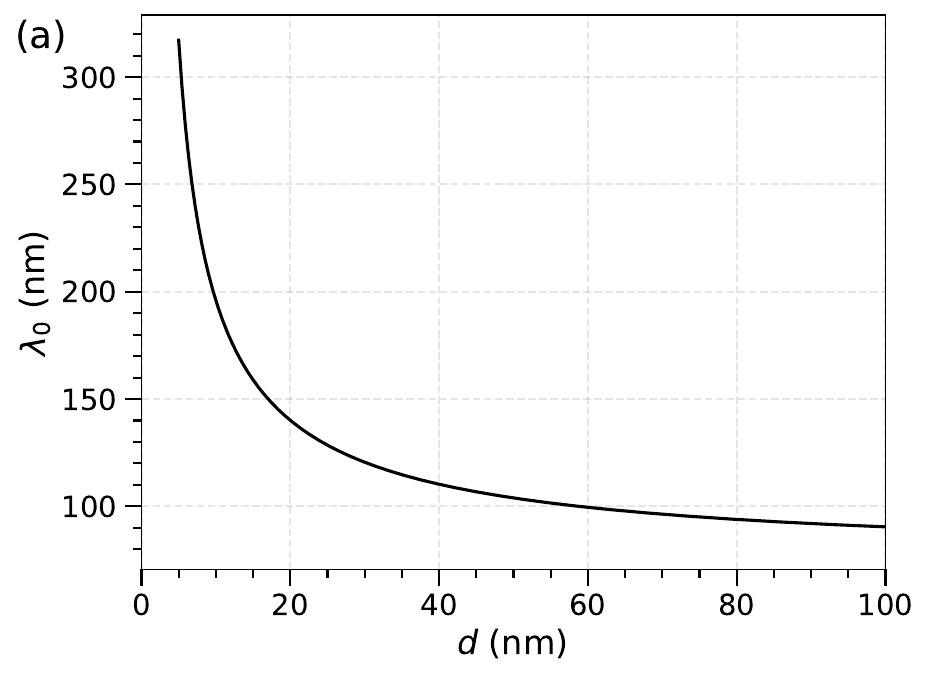}
\includegraphics[width=6cm]{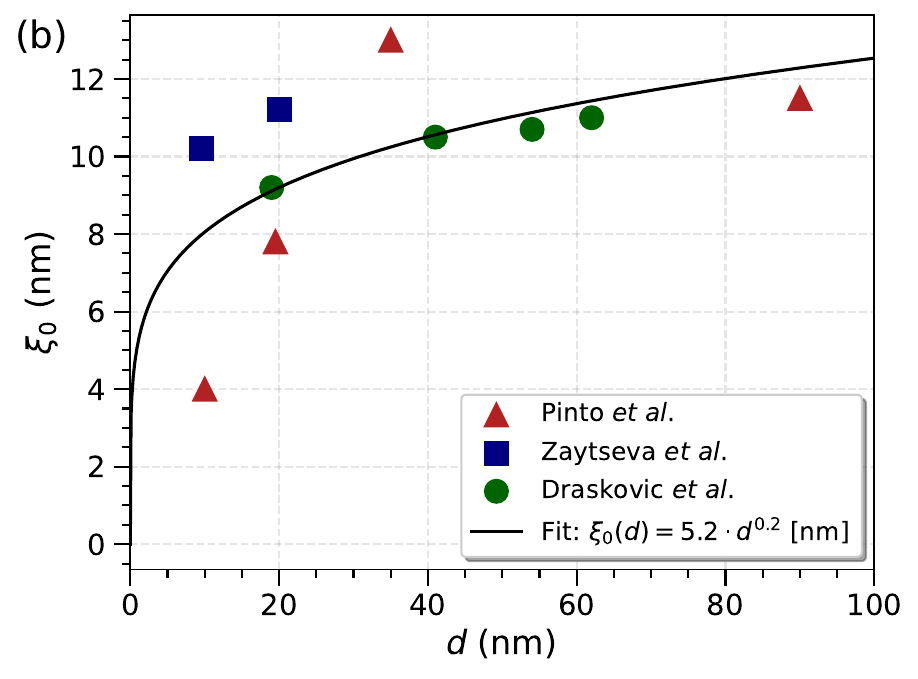}
\caption{The zero temperature superconducting length scales as a function of Nb film thickness on silicon substrate: (a) The magnetic penetration depth according to the empirical model provided by Ilin~\textit{et al.} \cite{Ilin2004peculiarities}. (b) The superconducting coherence length estimated by fitting a function $\xi_0(d)=k\cdot d^n$ to the experimental data provided by Pinto~\textit{et al.}~\cite{Pinto2018dimensional}, Zaytseva~\textit{et al.}~\cite{Zaytseva2020upper} and Draskovic~\textit{et al.}~\cite{Draskovic2013measuring}.   }
\label{lam_and_xi_vs_d_Fig-ab}
\end{center}
\end{figure*}

\noindent The choice of material is of paramount importance in the design of the device as the characteristic length scales of the superconductor ($\lambda_0$ and $\xi_0$) ultimately determine the dimensions of the device. Most prominently, this comes down to the $\xi_0$ which determines the radius of the vortex core. While radius of the pinning center in our device must be sufficiently large for it to efficiently pin a vortex, it must simultaneously fulfill $R_\mathrm{pin} \leq \xi_0$ so that vortices get attracted to it instead of experiencing repulsive surface potential. On the other hand, the dimensions of the device are limited by the spatial resolution of standard electron-beam lithography (EBL). To date, standard EBL devices can achieve 5\,nm resolution \cite{Sharma2022evolution, Yamazaki20045, Saifullah2022patterning}, which sets the lower limit for the dimensions of our design. That is, the $\xi$ of the superconducting material must be greater than equal to $5\,\mathrm{nm}$.

While the $\xi$ of most type-II superconducting materials can even increase well above 10\,nm near their characteristic superconducting transition temperatures, operation of the device at elevated temperatures is not desirable due to thermal noise and the resulting (classical) depinning of the vortex from the pinning center. Thus, we will choose the superconducting material solely based on the zero temperature coherence length $\xi_0$. This requirement, unfortunately, excludes all known high-temperature superconducting materials from our consideration. The only known type-II superconducting thin film for which $\xi_0>5\,\mathrm{nm}$ is (dirty) Nb. For our advantage, Nb is widely studied and applied superconducting material and is typically grown on silicon substrates using dc magnetron sputtering. The effective $\lambda_0$ and $\xi_0$ are determined by the mean free path of the Cooper pairs in the superconducting lattice, which ultimately comes down to the crystalline quality of the film \cite{Miller1959penetration, Minhaj1994thickness}. Consequently, the values of $\lambda_0$ and $\xi_0$ can be efficiently tuned by varying the thickness of the film ($d$). 

We have addressed $\lambda_0(d)$ using the empirical model provided by Ilin~\textit{et al.} \cite{Ilin2004peculiarities}. The model is in accordance with dirty limit Ginzburg-Landau theory and shows a sharp increase in $\lambda_0$ from the bulk value of $\sim 75\,\mathrm{nm}$ when the thickness is decreased below $d<40\,\mathrm{nm}$ as illustrated in Fig.~\ref{lam_and_xi_vs_d_Fig-ab}(a). In the case of $\xi_0(d)$ we faced a lack of reliable empirical models and, consequently, addressed it by fitting a phenomenological function $\xi_0(d) = k\cdot d^n$ via parameters $k$ and $n$ to available experimental data provided by Pinto~\textit{et al.}~\cite{Pinto2018dimensional}, Zaytseva~\textit{et al.}~\cite{Zaytseva2020upper} and Draskovic~\textit{et al.}~\cite{Draskovic2013measuring}. The fit, resulting in $\xi_0(d) = 5.2\cdot d^{0.2}\,[\mathrm{nm}]$, is illustrated in Fig.~\ref{lam_and_xi_vs_d_Fig-ab}(b). While the $\xi_0$ increases sharply when the film thickness is increased up to $20\,\mathrm{nm}$, most likely due to Si-Nb interface effects \cite{Pinto2018dimensional}, its value saturates between 10--15\,nm at larger thicknesses. 

The thickness dependent $\xi_0$, in particular, of dirty Nb films are desirable for the proposed device due to $\xi_0>5\,\mathrm{nm}$ for practically achievable film thicknesses ($d>1\,\mathrm{nm}$). This enables one to achieve strong pinning via an EBL generated artificial pinning center with dimensions comparable to $\xi_0$. Thus, we will fix the material of the device to Nb (on Si substrate).

\subsection{Dimensions}
\label{dimensions_section}
\noindent As concluded above, the standard EBL machines can achieve 5\,nm spatial resolution, setting the ultimate lower limit for the device dimensions and, in particular, for the $R_\mathrm{pin}$. To achieve strong pinning we want to match $R_\mathrm{pin}\approx \xi_o(d)$. Looking at the $\xi_0(d)$ in Fig.~\ref{lam_and_xi_vs_d_Fig-ab}(b), this condition can be easily met. However, since $m_\mathrm{d}\propto d$ we want to have as small $d$ as possible in order to minimize the mass of the vortex. This is because higher masses result in more localized quantum wave functions and decreased gap between the associated energy eigenstates, making the characterization of the device challenging. With this in mind, we will fix the film thickness to $d=5\,\mathrm{nm}$, resulting in $\xi_0(d=5\,\mathrm{nm})\approx7\,\mathrm{nm}$. In consequence, we will fix the radius of the EBL generated pinning center to $R_\mathrm{pin}=7\,\mathrm{nm}$.

With the fixed $d=5\,\mathrm{nm}$ and $R_\mathrm{pin}=7\,\mathrm{nm}$, we are left with determining the size of the superconducting island $R_\mathrm{dev}$ together with the applied magnetic field $B$ required for single vortex entry and subsequent pinning in our device. Both of these parameters affect the energy barrier (surface barrier) for vortex entry (or exit) in the device \cite{Schweigert1999flux, Berdiyorov2005surface}. The surface barrier together with vortex-vortex interactions determine the steady vortex state of the device, potentially including both pinned and unpinned vortices along with giant vortex states \cite{Zha2006superconducting}. In order to find combinations of $R_\mathrm{dev}$ and $B$ that allow single vortex pinning in our device, we have studied the associated steady vortex states using time-dependent Ginzburg-Landau (TDGL) simulations. 

\begin{figure*}[t!]
\begin{center}
\includegraphics[width=13.0cm]{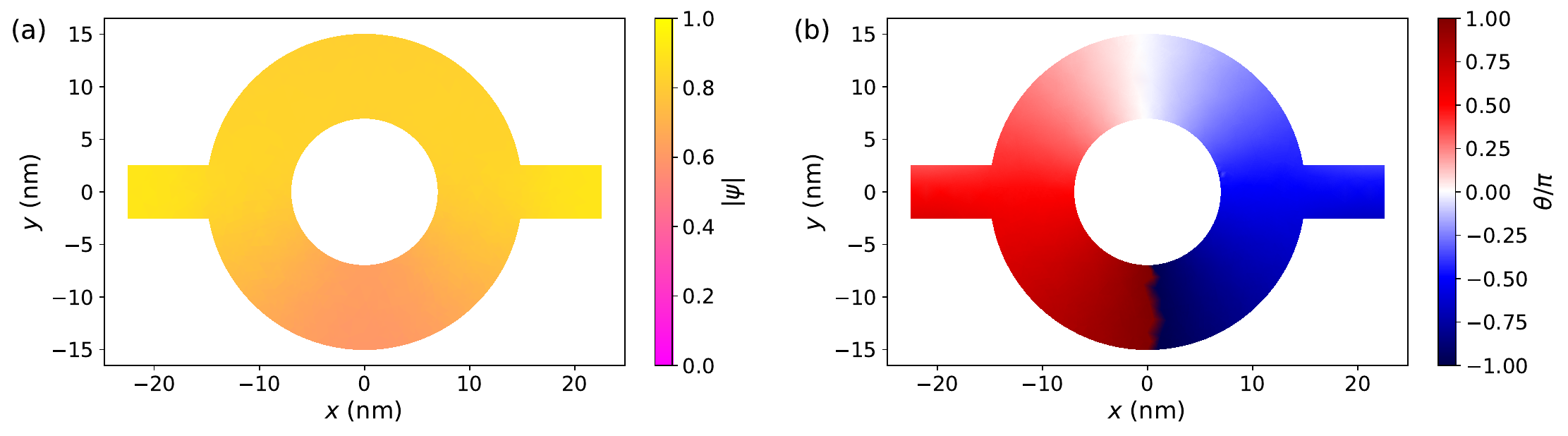}
\caption{The results of the time-dependent Ginzburg-Landau simulations for the optimized device ($R_\mathrm{dev}=15\,\mathrm{nm}$, $R_\mathrm{pin}=7\,\mathrm{nm}$, $d=5\,\mathrm{nm}$) at $B=6\,\mathrm{T}$ after 500 time steps, resulting in a steady state where a single vortex is trapped in the pinning center. (a) The magnitude of the order parameter and (b) the phase of the order parameter across the device.}
\label{TDGL_simulation_results_ab}
\end{center}
\end{figure*}

The TDGL simulations were performed using the open source \textit{pyTDGL} framework developed by Bishop--Van Horn~\cite{Van_Horn2023}. The Nb based device was modeled using the empirical formulas for $\lambda(d)$ and $\xi(d)$ discussed in section~\ref{material_section} (see Fig.~\ref{lam_and_xi_vs_d_Fig-ab}(a)--(b)). We have set the thickness-independent parameter $u=5.79$ based on Refs.~\cite{Van_Horn2023, Berdiyorov2009kinematic}, characterizing the ratio of the relaxation times for the order parameter, amplitude and phase. Additionally, we have fixed the parameter $\gamma = 20$, characterizing the strength of inelastic electron-phonon scattering, similarly to Refs.~\cite{Berdiyorov2009kinematic, Xue2024case}. A total of 28 distinct combinations of $R_\mathrm{dev}\in [15,\,20,\,...,\,30]\,\mathrm{nm}$ and $B \in [2,\,2.5,\,...,\,5]\,\mathrm{T}$ were simulated for a total of 400 time-steps. The presence of vortices within the structure of the device can be identified from the phase winding around a closed loop encircling its center, where single pinned vortex is associated by $\Delta\theta=2\pi$ as clearly indicated in Fig.~\ref{TDGL_simulation_results_ab}(b).

We observed that the applied $B$ field required for vortex entry decreased as a function of $R_\mathrm{dev}$. In the case of the smallest considered $R_\mathrm{dev}$ of $15\,\mathrm{nm}$ vortex entry was not even observed at the maximum field of 5\,T. We later manually evaluated it to be around 6\,T, above which single vortex pinning dominated at least until 8\,T, that was the highest field used in the simulations. Meanwhile, for all other devices, with $R_\mathrm{dev}>30\,\mathrm{nm}$, multiple vortices started penetrating the film for $B\geq5\,\mathrm{T}$ and giant vortices were also observed. It should be pointed out that the observed penetration of individual vortices through the outer edges of the device clearly distinguishes the considered system from fluxons in mesoscopic superconducting rings \cite{Petkovic2016deterministic}. 

For our consideration the optimum value for $R_\mathrm{dev}$ is the smallest considered radius of $15\,\mathrm{nm}$ due to the observed resilient single vortex pinning in a wide field range $6\,\mathrm{T}\leq B \leq 8\,\mathrm{T}$. The smallest plausible field of $B=6\,\mathrm{T}$ is below the zero temperature dirty limit upper critical field $B_{\mathrm{c}2} = \Phi_0/(2\pi \xi_0^2)$, where the empirical $\xi_0(d=5\,\mathrm{nm})\approx7\,\mathrm{nm}$ (Fig.~\ref{lam_and_xi_vs_d_Fig-ab}(b)) results in $B_{\mathrm{c2}}\approx 6.7\,\mathrm{T}$, also aligning with experimental observations \cite{Bose2006upper}.

To summarize, we have decided to fix the thickness of the Nb film to $d=5\,\mathrm{nm}$ together with the pinning site radius $R_\mathrm{pin}=7\,\mathrm{nm}$ corresponding to the associated $\xi_0$ (Fig.~\ref{lam_and_xi_vs_d_Fig-ab}(b)). Time-dependent Ginzburg-Landau simulations suggest that resilient single vortex pinning occurs for superconducting island of radius $R_\mathrm{dev}=15\,\mathrm{nm}$ when applied with $B=6\,\mathrm{T}$ field. The magnitude and phase of the simulated superconducting order parameter in the single vortex steady state is visualized in Fig.~\ref{TDGL_simulation_results_ab}(a)--(b). In the following sections we will address the dynamics of the pinned vortex in this particular device.

\section{Pinned Abrikosov Vortex}
\label{Pinned_Abrikosov_vortices_section}
\subsection{Pinning Potential}
\label{pinning_potential_section}
\noindent We model the pinning potential of the device with our previously used approach for high-temperature superconducting films in Refs.~\cite{Rivasto2025A, Aye2024enhanced}. That is, by relying on the variational solution of the superconducting order parameter ($\psi$) in the vicinity of vortex core, given by \cite{Clem1975simple}
\begin{equation}
    \label{vortex_core_Eq}
    \psi_\mathrm{v}(r) =   \frac{r}{\left( r^2 + 2\cdot \xi(T)^2 \right)^{1/2}}  \cdot \mathrm{e}^{i\theta},
\end{equation}
where $r=\sqrt{x^2+y^2}$ is the radial distance from cylindrically symmetric vortex core about the $z$-axis and $\theta$ is the spatially varying phase of $\psi$. The shape of the pinning potential between the Abrikosov vortex (Eq.~(\ref{vortex_core_Eq})) and the pinning center in our device can be calculated with the general formula 
\begin{equation}
\label{general_pinning_potential_Eq}
    u(r_\mathrm{v}) \propto \int_{-\infty}^{+\infty} \chi(r)\cdot \left|\psi_\mathrm{v}(r-r_\mathrm{v}) \right|^2 dr,
\end{equation} 
where the dimensionless \textit{form function} $\chi(r)=1-|\psi(r)|^2$ encodes the information about the spatial variation of the superconducting order parameter near the pinning center. For a pinning center of radius $R$ located at the origin, the form function is given by 
\begin{equation}
\label{Chi_eq}
\chi(r)=
\begin{cases}
1, & \text{if $|r|\leq R$}\\
1 - \tanh^2\left( \frac{|r-R|}{ \sqrt{2}\cdot \xi(T)} \right),& \text{otherwise},
\end{cases}
\end{equation} 
where the latter term takes into account the temperature dependent regeneration of $\psi$ from 0 to 1 as a function of distance from the edge from the pinning center. 

While the shape of the pinning potential can be numerically solved from Eqs.~(\ref{vortex_core_Eq})--(\ref{Chi_eq}), the obtained potential must be min-max normalized (between -1 and 0) and then scaled by the absolute pinning energy ($u_0$). We have previously concluded in Ref.~\cite{Rivasto2025A} that the pinning energy (per unit film thickness) calculated using the formula \cite{Blatter1994vortices, Nelson1993boson, Klaassen2001vortex}
\begin{equation}
\label{theoretical_pinning_energy_Eq}
    u_0 = -\frac{\Phi_0^2}{8\pi \mu_0 \lambda^2} \ln\left( 1 + \frac{R^2}{2\xi^2} \right)
\end{equation}
aligns well with experimentally determined values for large columnar defects in {Y}{B}a$_2${C}u$_3${O}$_{7-\delta}$ films. With the lack of a better model, we will adopt this formula for estimating the pinning energy in our device.

\begin{figure*}[t!]
\begin{center}
\includegraphics[width=6.0cm]{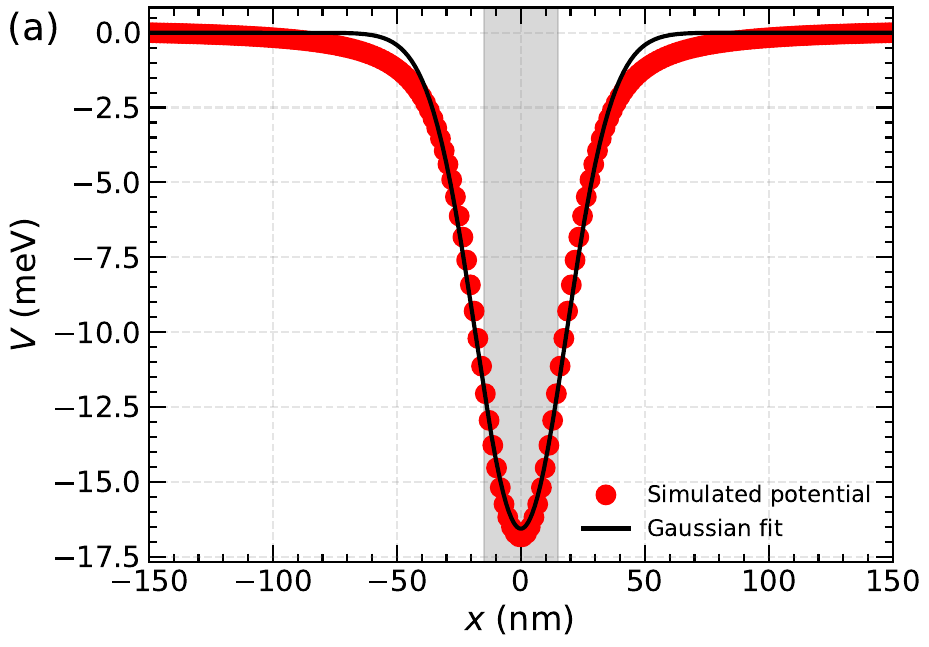}
\includegraphics[width=6.0cm]{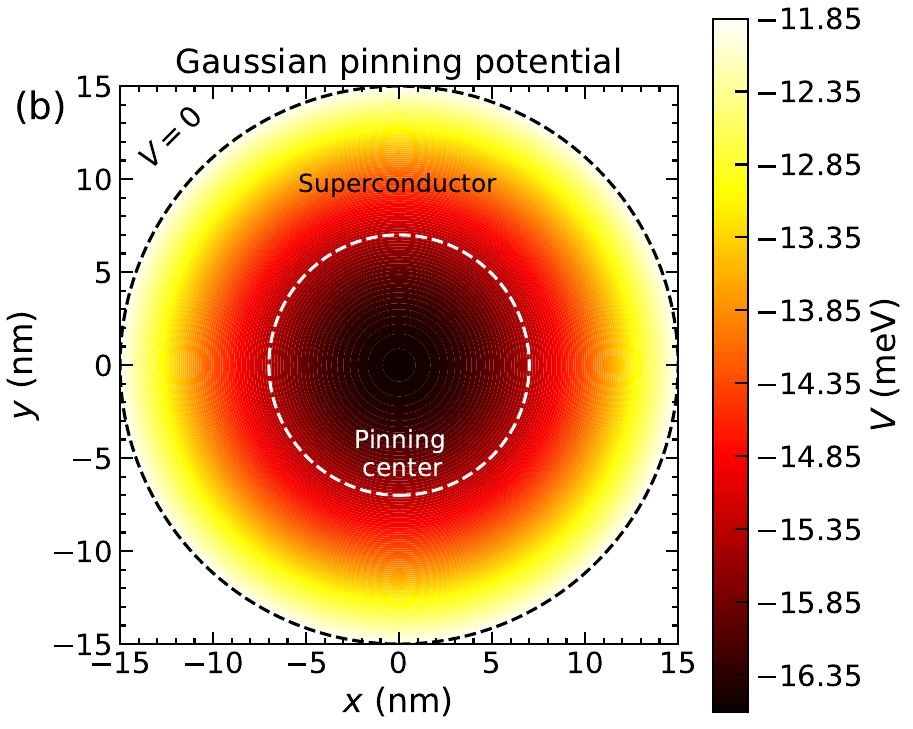}
\caption{(a) The numerically calculated one-dimensional pinning potential based on Eqs.~(\ref{vortex_core_Eq})--(\ref{Chi_eq}) (where $r=\sqrt{x^2+y^2}$ is expressed as $r=x$ at $y=0$) and a fit of a Gaussian potential $V(x)=a\cdot \mathrm{exp}(-\gamma x^2)$ via parameters $a$ and $\gamma$ to the numerical data. The fit resulted $a=-15.57$ and $\gamma=0.0015$. The gray region represents the scale of the device, corresponding to $x\in [-R_\mathrm{dev},\,R\mathrm{_\mathrm{dev}}]$. (b) The resulting cylindrically symmetric two-dimensional Gaussian potential based on the previous fit, that is used for solving the Schrödinger equation. The potential is set to $V=0$ when the distance from the origin is greater than $R_\mathrm{dev}=15\,\mathrm{nm}$. }
\label{Gaussian_fit_fig}
\end{center}
\end{figure*}

\subsection{Mass of the Abrikosov Vortex}
\label{mass_of_abrikosov_vortex_section}
\noindent Modeling of the pinned Abrikosov vortex as a quasiparticle, either in classical or quantum domain, requires one to assign it with an effective mass ($m_\mathrm{v}$). Determining the $m_\mathrm{v}$, or more generally $m_\mathrm{v}$ per unit film thickness ($\mu_\mathrm{v}$), has been widely debated in the past with different theoretical and experimental approaches (see extensive review by Rusakov~\textit{et al.} \cite{Rusakov2017oscillations}). Several different effects have been proposed to contribute to the effective mass of a vortex. These include i) the mass of quasiparticles within the vortex core \cite{Suhl1965inertial, Kopnin1998dynamic}, ii) classical dynamic interaction of vortex core with rotating superfluid electron liquid \cite{Baym1983hydrodynamics}, vortex motion induced dipolar electric field \cite{Coffey1991dipolar} and vortex motion induced elastic deformation of surrounding crystal lattice \cite{Coffey1994deformable}. From these, the quasiparticle mass within the core of the vortex is believed to dominate. However, there does not exist a single general model or experimental approach for addressing vortex masses in superconducting thin films. In consequence, we will rely on studying the Abrikosov vortex quasiparticle dynamics using a range of $m_\mathrm{v}$s that have been proposed by recent experiments. 

Closely related to our interest for the Nb based device, Golubchik~\textit{et al.} \cite{Golubchik2012mass} have measured vortex mass per unit film thickness in Nb film by measuring the motion of individual vortices using high-resolution magneto-optical imaging together with ultrafast heating and cooling technique. They evaluated the effective quasiparticle vortex mass to lie between $10^9\,m_\mathrm{e}/\mathrm{m}<\mu_\mathrm{v}<10^{11}\,m_\mathrm{e}/\mathrm{m}$, where $m_\mathrm{e}$ is the mass of an electron. We also want to point out the recent terahertz frequency magnetic circular dichroism measurements by Tesarv~\textit{et al.}~\cite{Tesavr2021mass, Tesavr2024perspective}, who have estimated  a vortex mass of $\mu_\mathrm{v} \sim 10^{10}\,m_\mathrm{e}/\mathrm{m}$ for {Y}{B}a$_2${C}u$_3${O}$_{7-\delta}$ films at 45\,K temperature. Despite being different material, the estimated mass aligns well with the range proposed by Golubchik. 

Regarding the experimental measurement of the quantum state of a pinned Abrikosov vortex, it is natural to hope that the vortex has a small mass, since the resulting energy eigenstates are then more widely separated and therefore more easily measurable. With this in mind, we conservatively restrict the range of vortex masses under consideration to between $10^{10}\,m_\mathrm{e}/\mathrm{m}$ and $10^{13}\,m_\mathrm{e}/\mathrm{m}$, spanning three orders of magnitude. For our device, with the film thickness fixed at $d = 5\,\mathrm{nm}$, this range corresponds to $m_\mathrm{v} \in [10m_\mathrm{e},\,10^4m_\mathrm{e}]$.

\section{Quantization}
\label{quantization_section}
\noindent In this section, we will solve the two-dimensional time-independent Schrödinger equation $\hat{H}\psi(x,y)=E\psi(x,y)$ for the pinned Abrikosov vortex in the optimized design of the device (Fig.~\ref{TDGL_simulation_results_ab}). The Hamiltonian for the system is simply
\begin{align}
    \hat{H} = -\frac{\hbar^2}{2m} \left( \frac{\partial^2 }{\partial x^2} + \frac{\partial^2 }{\partial y^2}\right) + V(x,y),
\end{align}
where $V(x,y)$ is the pinning potential calculated by Eqs.~(\ref{vortex_core_Eq})--(\ref{Chi_eq}) introduced in section \ref{pinning_potential_section}. In order to mitigate issues arising from numerical instability in solving the eigenfunctions (vectors) and eigenvalues of the Hamiltonian using the finite difference method (FDM), we have approximated $V(x,y)$ resulting from Eqs.~(\ref{vortex_core_Eq})--(\ref{Chi_eq}) with an analytical function. To do this, we have fitted a Gaussian function $G(x)= a \cdot \mathrm{exp}(-\gamma x^2)$ to the numerically calculated potential. The fit is illustrated in Fig.~\ref{Gaussian_fit_fig}(a), resulting in $a=-16.57\,\mathrm{meV}$, $\gamma=0.0015\,\mathrm{nm}^{-2}$. The fitted Gaussian can be observed to describe the pinning potential well within the dimensions of the device ($R_\mathrm{dev}=15\,\mathrm{nm}$). We have also tried fitting Lorentzian and parabolic functions but the Gaussian fit results in lowest sum of residuals, in particular near the bottom of the potential well. We have used the fitted Gaussian to create the 2-dimensional cylindrically symmetric pinning potential that is truncated to $V(r)=0$ for $r=\sqrt{x^2+y^2}>R_\mathrm{dev}$. The resulting potential is illustrated in Fig.~\ref{Gaussian_fit_fig}(b). The size of the grid used in the FDM calculations ranged from $x\in [-15\,\mathrm{nm},\,15\,\mathrm{nm}]$ and $y\in [-15\,\mathrm{nm},\,15\,\mathrm{nm}]$, exactly as depicted in Fig.~\ref{Gaussian_fit_fig}(b). The calculations were performed using 500 grid points in both the $x$ and $y$ directions, corresponding to a grid spacing of $\Delta x = \Delta y = 0.06\,\mathrm{nm}$. Increasing the grid density above this yielded only negligible deviations in the results. Dirichlet boundary conditions were applied, such that the wavefunction vanishes at the domain boundaries.

\begin{figure*}[t!]
\begin{center}
\includegraphics[width=13cm]{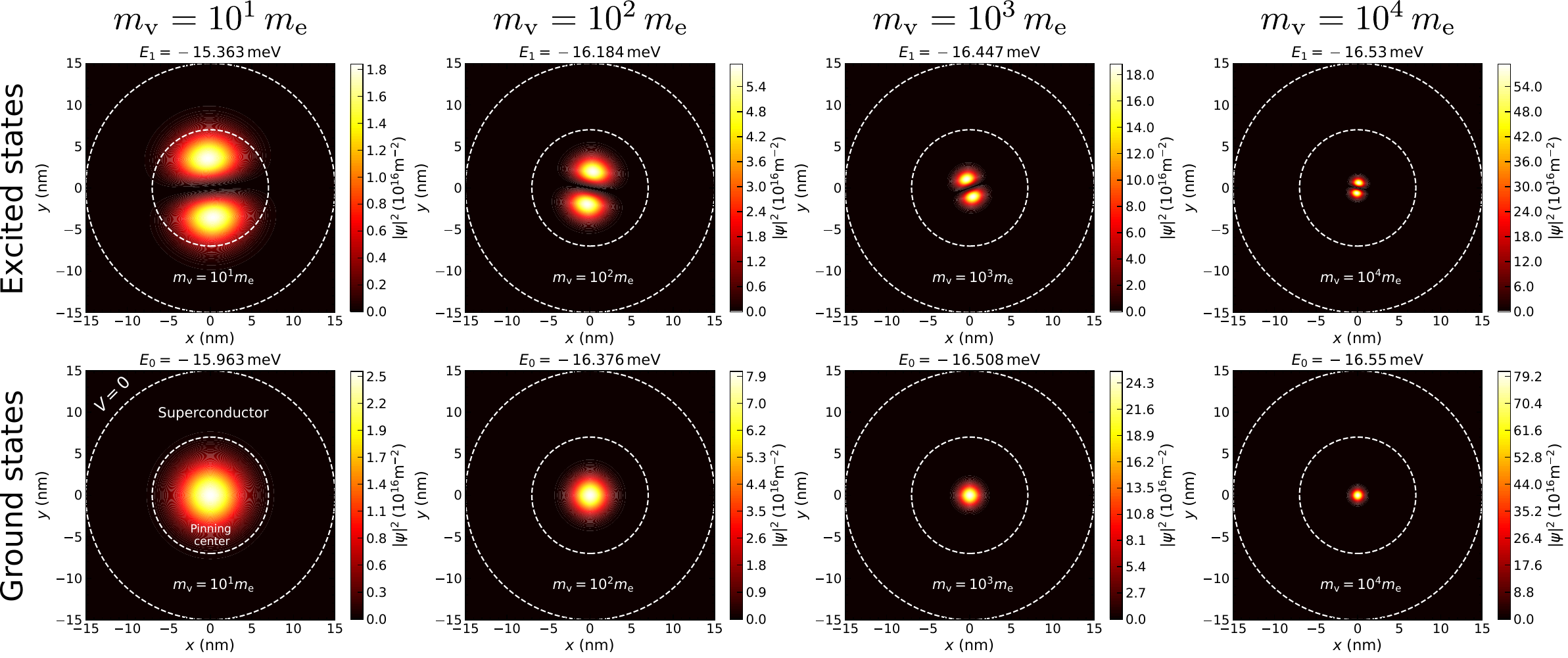}
\caption{Visualizations of the ground state and the first excited state wavefunctions (continuous normalization) of the pinned vortex for different values of $m_\mathrm{v}$. The data is obtained from the numerical solutions of the time-independent Schrödinger equation for the Gaussian pinning potential illustrated in Fig.~\ref{Gaussian_fit_fig}(b). The dashed white regions represent the dimensions of the pinning center ($R_\mathrm{pin}=7\,\mathrm{nm}$) and device ($R_\mathrm{dev}=15\,\mathrm{nm}$). }
\label{eigenfunctions_fig}
\end{center}
\end{figure*}

\begin{figure}[t!]
\begin{center}
\includegraphics[width=6cm]{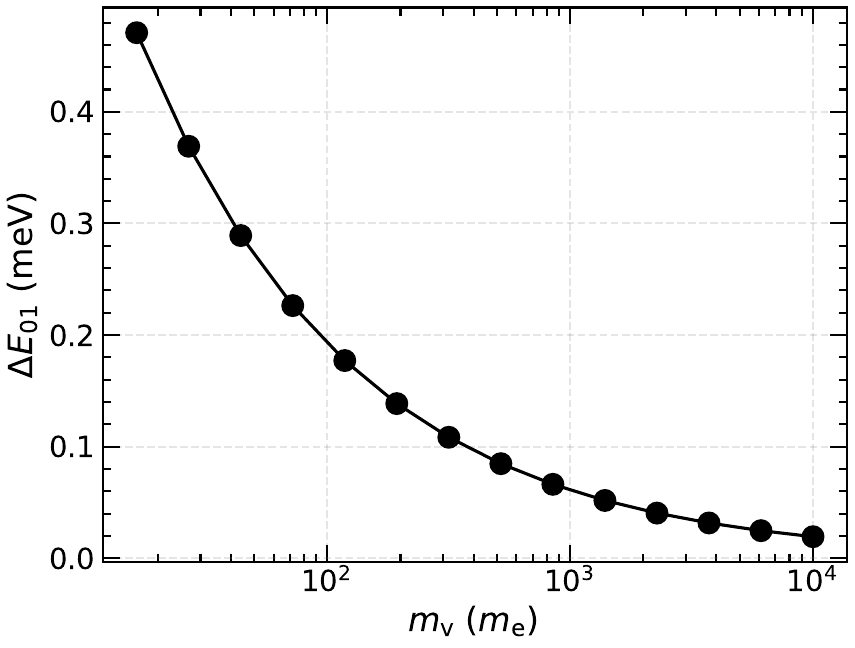}
\caption{The excitation energy between the ground state and first excited state of the pinned vortex as a function of its mass. }
\label{de-mv_fig}
\end{center}
\end{figure}

The Gaussian pinning potential of Fig.~\ref{Gaussian_fit_fig}(b) is then used to solve the Schrödinger equation for different values of vortex masses within the empirically motivated range $m_\mathrm{v} \in [10 m_\mathrm{e},\,10^{4} m_\mathrm{e}]$ (see sec.~\ref{mass_of_abrikosov_vortex_section}). The eigenfunctions of the ground state and the first excited state for different vortex masses are illustrated in Fig.~\ref{eigenfunctions_fig}. As expected, the eigenfunctions become more localized as the vortex mass is increased. This is accompanied by decrease in the excitation energy between the ground and excited state ($\Delta E=E_1 - E_0$). The decrease of excitation energy as a function of vortex mass is illustrated in Fig.~\ref{de-mv_fig}. Even for vortex masses on the order of $10^4 m_\mathrm{e}$, the excitation energy remains well above the $\mu\mathrm{eV}$ range, indicating promising experimental accessibility for determining the quantum state of the vortex.

\section{Measuring the Quantum State}
\label{measuring_quantum_state_section}
\begin{figure*}[t!]
\begin{center}
\includegraphics[width=6cm]{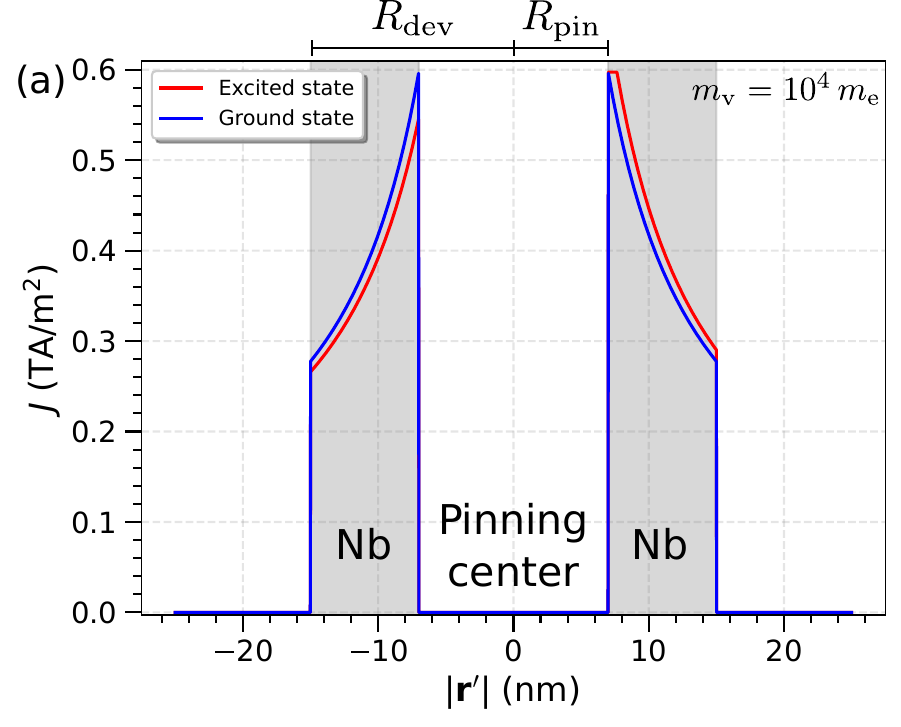}
\includegraphics[width=6cm]{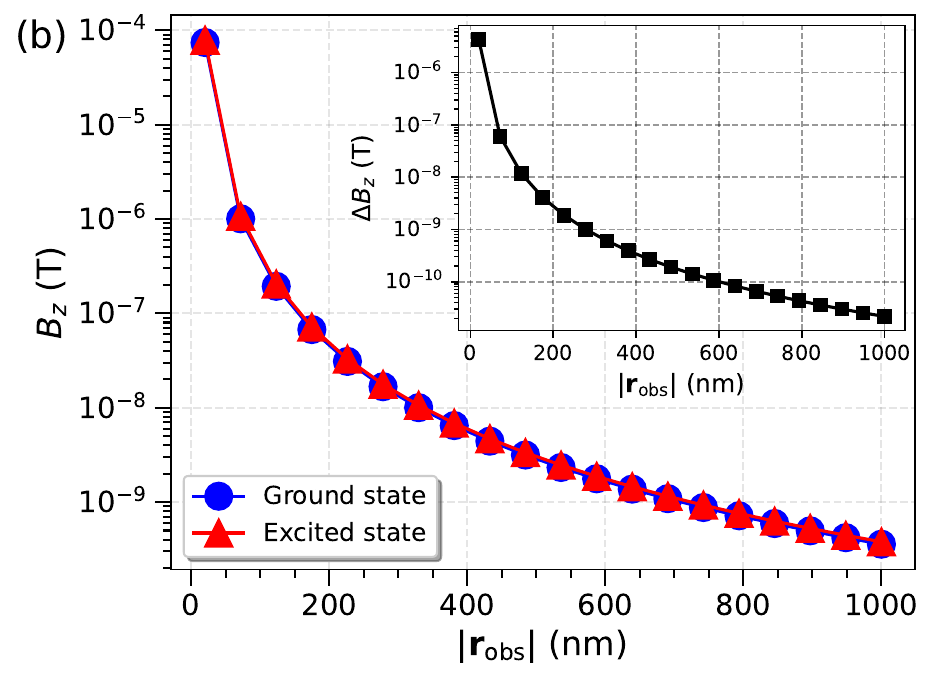}
\caption{(a) The current distributions calculated from Eq.~(\ref{J_distribution}) in the cross-section of the optimized device (see Fig.~\ref{device_general_illustration_fig}) for the ground and the first excited states of the pinned Abrikosov vortex with $m_\mathrm{v}=10^4\,m_\mathrm{e}$. The center of the vortex is at the origin for ground state, but is shifted by $\Delta s \approx 0.65\,\mathrm{nm}$ along the positive $x$-axis for the excited state according to Fig.~\ref{eigenfunctions_fig}. (b) The decay of magnetic fields as a function of radial distance from the outer edge of the (cylindrically symmetric) device ($|\mathbf{r}_\mathrm{obs}|>R_\mathrm{dev}$) resulting from the aforementioned current distribution. The inset shows the difference of $B_z(|\mathbf{r}_\mathrm{obs}|)$ between the ground and excited states. }
\label{biot_savart_results_fig_ab}
\end{center}
\end{figure*}
\noindent Next, we discuss the possibilities for experimental distinction between the ground and the first excited state of the pinned vortex. This is ultimately done by measuring the spatially dependent magnetic field generated by the vortex. Here, we will calculate the difference of the magnetic field profiles of a vortex between its ground and excited states using the Biot-Savart law, where the circulating supercurrent of the vortex is the source of the magnetic field far away from the device.

The Biot-Savart law can be expressed as \cite{Griffiths2023introduction}
\begin{align}
\label{biot-savart_eq}
    \int d\mathbf{B} = \frac{\mu_0}{4\pi} \int \frac{ I(\mathbf{r'}) d\mathbf{l} \times (\mathbf{r}_\mathrm{obs}-\mathbf{r'}) }{|\mathbf{r}_\mathrm{obs}-\mathbf{r'}|^3},
\end{align}
where $\mathbf{r}_\mathrm{obs}$ is the observation position at which $\mathbf{B}$ is calculated and $\mathbf{r}'$ is the position of the source associated with a current 
\begin{align}
\label{I_eq}
    I(\mathbf{r'})= \int J(\mathbf{r'}) \cdot d\mathbf{A}.
\end{align}
Considering cylindrically symmetric current distribution, Eq.~(\ref{I_eq}) can be reduced to $I(\mathbf{r'}) = J(\mathbf{r'})\cdot d\cdot dr$, where $d=5\,\mathrm{nm}$ is the film thickness and $dr$ is the differential radial displacement vector in cylindrical coordinates. We model $J(\mathbf{r'})$ as \cite{Poole2014superconductivity}
\begin{equation}
\label{J_distribution}
J(\mathbf{r'})=\frac{\Phi_0}{2\pi \mu_0 \lambda^3} \cdot 
\begin{cases}
0,\text{ for $0 \leq |\mathbf{r'}| \leq R_\mathrm{pin}$ or $|\mathbf{r'}|>R_\mathrm{dev}$ } \\
\mathrm{K_1\left( \frac{\xi}{\lambda} \right)},\text{ for $|\mathbf{r'}-\mathbf{r}_\mathrm{core}|\leq \xi$} \\
\mathrm{K_1\left( \frac{|\mathbf{r'}-\mathbf{r}_\mathrm{core}|}{\lambda} \right)},\text{ for $|\mathbf{r'}-\mathbf{r}_\mathrm{core}|\geq \xi$}
\end{cases}
\end{equation}
where $\mathrm{K_1}$ is first-order modified Bessel function and $\mathbf{r}_\mathrm{core}$ is the position of the vortex core. For the ground state $\mathbf{r}_\mathrm{core}$ is (on average) at the origin, while for the excited state it shifts away from the origin by amount $|\mathbf{r}_\mathrm{core}|=\Delta s$ (see Fig.~\ref{eigenfunctions_fig}). This change in the average position of the vortex will change the magnetic field profile measured in the observation point $\mathbf{r}_\mathrm{obs}$, that we consider the main measurable quantity regarding the determination of the quantum state of the vortex. As seen from the solved eigenfunctions presented in Fig.~\ref{eigenfunctions_fig}, the $\Delta s$ is strongly dependent on $m_\mathrm{v}$. Here, we will only consider the most difficult case regarding distinguishing between the ground and excited states of the vortex, that is with the highest considered vortex mass $m_\mathrm{v}=10^4\,m_\mathrm{e}$ associated with the smallest $\Delta s$. We have evaluated $\Delta s \approx 0.65\,\mathrm{nm}$ directly from Fig.~\ref{eigenfunctions_fig}. The resulting differences in the current distributions across the cross section of the device are illustrated in Fig.~\ref{biot_savart_results_fig_ab}(a). We have used these cylindrically symmetric current distributions to solve the radial decay of the magnetic field along the $z$-direction ($B_z$) using the Biot-Savart law (Eq.~(\ref{biot-savart_eq})). The results are presented in Fig.~\ref{biot_savart_results_fig_ab}(b), the inset of which shows the difference between the magnetic field generated by a vortex in ground and excited states at the corresponding position. The difference in the magnetic field decays exponentially, but remains well above $\mathrm{nT}$ range for $|\mathbf{r}_\mathrm{obs}| \lesssim 300\,\mathrm{nm}$. We conclude that sensing the state of the vortex via the calculated (worst case) subtle change in its magnetic field is experimentally plausible. For example, nitrogen vacancy (NV) center based magnetometers are mature and widely used technology being able to reach $\mathrm{pT}/\sqrt{\mathrm{Hz}}$ sensitivity with nm-scale spatial resolution \cite{Barry2020sensitivity, Kenny2025quantum}. There also exists a variety of chip-scale atomic magnetometers with $\mathrm{fT}/\sqrt{\mathrm{Hz}}$ sensitivity \cite{Kitching2018chip}.

\section{Conclusions}
\label{conclusions_section}
\noindent This study investigated the possibility of observing quantized energy states of a single, strongly pinned Abrikosov vortex. In summary, we have presented a detailed design process for a Nb-based superconducting device capable of robustly trapping a single Abrikosov vortex within its structure under a magnetic field of 6\,T. The proposed device consists of a cylindrically symmetric Nb film, 5\,nm thick and 30\,nm in diameter, featuring a 14\,nm diameter central cavity that serves as a pinning site.

We have calculated the energy levels and corresponding quantum wavefunctions of the pinned vortex by numerically solving the two-dimensional time-independent Schrödinger equation for the system, considering a broad yet conservative range of vortex masses reported in the literature. The energy difference between the ground state and the first excited state is found to exceed the $\mu$eV range even for the highest mass estimates, suggesting that these states could be experimentally accessible via microwave spectroscopy \cite{Pompeo2008reliable, Alimenti2020microwave}, which we consider the most important step toward the experimental realization of these phenomena. Furthermore, we have shown that it is plausible to probe the quantum state of the vortex indirectly by detecting spatial variations in the magnetic field generated by the circulating supercurrents, although this approach is technically challenging.

The main limitation of this study lies in the assumption of negligible dissipation for the pinned Abrikosov vortex. The dynamics of Abrikosov vortices in homogeneous superconducting films are indeed known to be highly viscous due to quasiparticle excitations in the vortex core \cite{Bardeen1965theory}. However, in the proposed device, the vortex core resides entirely within the nanofabricated columnar pinning site, which is expected to strongly suppress viscous damping arising from quasiparticle excitations. Although a theoretical estimate of the residual viscous damping and its influence on the quantum dynamics remains challenging, experimental realization of the proposed device would provide valuable opportunities to probe these effects directly.

A second limitation concerns the assignment of the vortex mass. Since the vortex mass is believed to originate predominantly from quasiparticles within the vortex core, this contribution may also be significantly suppressed in the considered configuration. This suppression would, in fact, favor the experimental observation of vortex quantum states, as the calculated energy spacing between the ground and first excited states increases with decreasing vortex mass. Thus, the excitation energies obtained in this work can be regarded as conservative lower bounds.

In conclusion, we have provided strong theoretical evidence supporting the experimental measurability of quantized energy states in a pinned Abrikosov vortex. The experimental realization of the proposed device would, for the first time, enable direct observation of quantum features associated with Abrikosov vortices, along with a quantitative assessment of their viscous damping and effective mass in a strongly pinned regime. We further expect that the proposed device concept may find applications in cryogenic memory technologies \cite{Ortlepp2014access, Golod2015single, Aziz2025superconducting} and quantum computing architectures \cite{Gyongyosi2019survey, Alam2023cryogenic}.

\section{Acknowledgments}
E.O.R. acknowledges support from the European Research Council (ERC) under the European Union’s Horizon 2020 research and innovation program Grant agreement No. 948689 (AxionDM). E.O.R. wants to thank Pietro Butti for the insightful discussions and help regarding the numerical solution of the Schrödinger equation.

\bibliography{bibliography.bib}

\end{document}